\documentclass[twocolumn,preprintnumbers,nofootinbib]{revtex4}
\usepackage{graphicx}
\usepackage{amssymb}
\usepackage{color}
\usepackage{enumerate}
\usepackage{amsmath}
\def\beq{\begin{equation}}
\def\eeq{\end{equation}}
\def\beqn{\begin{eqnarray}}
\def\eeqn{\end{eqnarray}}

\def\con{\color{black}} 

\renewcommand{\texttt}{{}}
\newcommand{\be}{\begin{eqnarray}}
\newcommand{\ee}{\end{eqnarray}}

\begin{document}

\title{
{Finite quantum 
gauge theories
}}

\author{Leonardo Modesto$^1$}
\email{lmodesto@fudan.edu.cn}

\author{Marco Piva$^2$}
\email{piva0505@gmail.com}

\author{Les\l aw Rachwa\l{}$^1$}
\email{rachwal@fudan.edu.cn}

\affiliation{$^1$Center for Field Theory and Particle Physics and Department of Physics, 
Fudan University, 200433 Shanghai, China\\
$^2$Dipartimento di Fisica ``Enrico Fermi", Universit\`a di Pisa, 
Largo B. Pontecorvo 3, 56127 Pisa, Italy}

\date{\small\today}

\begin{abstract} \noindent
We explicitly compute the one-loop exact beta function for a 
nonlocal extension of the standard gauge theory, in particular Yang-Mills and QED. The theory, made of a weakly nonlocal
kinetic term and a local potential of the gauge field, 
is unitary (ghost-free) and perturbatively  super-renormalizable. Moreover, in the action we can always 
choose the potential (consisting of one ``killer operator") 
to make zero the beta function of the running gauge coupling constant.
The outcome is {\em a UV finite theory for any gauge interaction}. 
Our calculations are done in $D=4$, but the results can be generalized to even or odd spacetime dimensions. 
We compute the contribution to the beta function from two different killer operators by using two independent techniques, namely the Feynman diagrams and the Barvinsky-Vilkovisky traces.
By making the theories finite we are able to solve also the Landau pole problems, in particular in QED. 
Without any potential the beta function of the one-loop super-renormalizable theory shows a universal Landau pole in the running coupling constant in the ultraviolet regime (UV), regardless of the specific higher-derivative structure. However, the dressed 
propagator shows neither the Landau pole in the UV, nor the singularities 
 in the infrared regime (IR).  
\end{abstract}

\maketitle

We study a class of new actions of fundamental nature for gauge theories  
that are super-renormalizable or finite at quantum level. 
In particular we hereby present four physical objectives to be met 
in a finite theory of QED and in Yang-Mills gauge interactions: 
avoiding the Landau pole in QED or for the $U(1)$ sector of the standard model of particle physics (SM); 
having a better control over divergences in QCD; having more room for unification of the running coupling
constants in the super-renormalizable extension of the SM; stabilizing the Higgs potential. 
Moreover, whether we want to study gauge theories coupled to super-renormalizable or finite gravity, 
then the former have to possess the same quantum properties. 
Furthermore, 
scale-invariant gauge theories in $D=4$ can be promoted to 
conformally invariant ones. 
%
We also require the following two guiding principles to be common to all the
fundamental interactions: 
``{super-renormalizability or finiteness}" and ``validity of perturbative expansion" in the quantum field theory framework \cite{Anselmi-1}. 
The desired theories satisfy the following properties:
(i) gauge invariance; (ii) weak nonlocality (or quasi-polynomiality); (iii) unitarity; (iv) quantum super-renormalizability or finiteness.  
The main difference with quantum perturbative standard Yang-Mills theory (or Abelian quantum electrodynamics) lies in the second requirement, which makes possible to achieve 
unitarity and renormalizability at the same time in any spacetime dimension $D$. 

Next, by choosing a subclass of theories with sufficiently high number of derivatives in the UV, we may get even better control over perturbative divergences - we actually may get super-renormalizability. This means that infinities in the perturbative calculus appear only up to some finite loop order. Finally, by adding some operators, which are higher in powers of the gauge field strength, with specially adjusted coefficients we achieve finiteness, namely the beta function of gauge coupling can be consistently set to vanish. The outcome is a quantum theory for any gauge interaction free of any divergence at any order in the loop expansion,
and the problem of the Landau pole in the UV is solved. 
Moreover, by shifting the coefficients of the theory we can easily achieve asymptotic freedom (in the beta function) for all interactions, if this is desired for grand unification. 

In a different vein if the theory is one-loop super-renormalizable and with higher-derivatives, then in the beta function we inevitably find a Landau pole at high energy, because the beta function is universally negative. However, when looking at the dressed propagator of the theory (or the quantum effective action) we see that the behaviour in UV as well as in IR is without additional real poles and the interactions are suppressed at high energy. Indeed, in the UV it is the nonlocal higher-derivative 
operator that controls the high energy physics, whereas in IR the theory remains in the perturbative regime because of the universal negative 
sign of the beta function $\beta_\alpha$. To fix the notation we here 
define the divergent contribution to the effective action in dimension four to be 
$\Gamma_{\rm div} \equiv \frac{1}{\epsilon} \beta_{\alpha} \! \int \! d^4 x \,  {\rm tr} {\bf F}^2$, 
where $\alpha := 1/g^2$ and $g$ is the gauge coupling constant. 
\section{Nonlocal Gauge theories}
A consistent gauge-invariant theory for spin one massless particles 
regardless of the spacetime dimension fits in the following general class of theories 
\cite{modestoLeslaw}
\be
&&
\mathcal{L}_{\rm gauge} = - \frac{1}{4 g^2} {\rm tr}\con\left[   \,  {\bf F} \, e^{H({\cal D}^2_{\Lambda})} {\bf F} 
+  \mathcal{\bf V}_{ g \con} \con
\right]  
  . \label{gauge}
\ee
The theory above consists of a weakly nonlocal kinetic operator and a local curvature potential ${{\bf V}_{ g \con}
}\con$ crucial to achieve finiteness of the theory as we will show later. 
In (\ref{gauge}) the Lorentz indices and tensorial structures have been neglected. 
The notation on the flat spacetime reads as follows: we use the gauge-covariant box operator defined via ${\cal D}^2={\cal D}_\mu{\cal D}^\mu$, where ${\cal D}_\mu$ is a gauge-covariant derivative (in the adjoint representation) acting on gauge-covariant field strength ${\bf F}_{\rho\sigma} = F_{\rho\sigma}^a T^a$ of the gauge potential $A_{\mu}$ (where $T^a$ are the generators of the gauge group in the adjoint representation.) 
The metric tensor $g_{\mu \nu}$ has 
signature $(- + \dots +)$. 
 We employ the following definition, ${\cal D}^2_{\Lambda} \equiv  {\cal D}^2/\Lambda^2$, where $\Lambda$ is an invariant mass scale in our fundamental theory. 
Finally, the entire function  $V^{-1}(z) \equiv \exp H(z)$ ($z \equiv  {\cal D}^2_\Lambda$) in (\ref{gauge})
satisfies the following general conditions \cite{kuzmin}, \cite{Tombo}: 
(i) $V^{-1}(z)$ is real and positive on the real axis and it has no zeros on the 
whole complex plane $|z| < + \infty$. This requirement implies, that there are no 
gauge-invariant poles other than for the transverse and massless gluons.
(ii) $|V^{-1}(z)|$ has the same asymptotic behaviour along the real axis at $\pm \infty$.
(iii) There exists $\Theta\in(0,\pi/2)$ such that asymptotically
$
|V^{-1}(z)| \rightarrow | z |^{\gamma + \frac{D}{2} - 2}$, when 
 $|z|\rightarrow + \infty$ with
$\gamma\geqslant D/2$ ($D$ is even and $\gamma$ natural)
for complex values of $z$ in the conical regions $C$ defined by: 
$C = \{ z \, | \,\, - \Theta < {\rm arg} z < + \Theta \, ,  
\,\,  \pi - \Theta < {\rm arg} z < \pi + \Theta\}.$
This condition is necessary to achieve the maximum convergence of the theory in
the UV regime. (iv) The difference $V^{-1}(z)-V^{-1}_{\infty}(z)$ is such that on the real axis
\be
\lim\limits_{|z|\rightarrow\infty}\frac{V^{-1}(z)-V^{-1}_{\infty}(z)}{V^{-1}_{\infty}(z)}z^m=0,\qquad  {\rm for\,\, all}\quad m\in \mathbb{N},
\ee
where $V^{-1}_{\infty}(z)$ is the asymptotic behaviour of the form factor $V^{-1}(z)$. Property (iv) is crucial for the locality of counterterms. The entire function $H(z)$ must be chosen in such a way that $\exp H(z)$ tends to a polynomial $p(z)$ in UV hence leading to the same divergences as in higher-derivative theories.

An explicit example of weakly nonlocal form factor $e^{H(z)}$, that has the properties (i)-(iv) can be easily constructed following \cite{Tombo}, 
\be
\hspace{0.1cm}
 &&e^{H(z)}
= e^{\frac{1}{2} \left[ \Gamma \left(0, e^{-\gamma_E}p(z)^2 \right)+ \log \left( p(z)^2 \right) \right] }
\nonumber \\
&& \underset{z\in\mathbb{R}}{=}
\sqrt{ p(z)^2} 
\left( 
1+ \frac{e^{-e^{-\gamma_E}p(z)^2}}{2 e^{-\gamma_E}p(z)^2} 
+ \dots 
\right)
\label{VlimitB}
 , 
\ee
where 
 $\gamma_E \approx 0.577216$ is the Euler-Mascheroni constant and 
$
\Gamma(0,x) = \int_x^{+ \infty}  d t \, e^{-t} /t 
$ 
is the incomplete gamma function with its first argument vanishing.
The  polynomial $p(z)$ of degree $\gamma + (D-4)/2$ is such that $p(0)=0$, which gives the correct low energy limit of our theory coinciding with the standard two-derivative Yang-Mills theory. In this case the 
$\Theta$-angle defining cones $C$ turns out to be $ \pi/ (4\gamma + 2(D-4))$.

The theories described by the action in (\ref{gauge}) are unitary and perturbatively renormalizable at quantum level in any dimension as we are going to explicitly show in the following subsections. 

Moreover, at classical level many evidences endorse that we are dealing with ``{\em gauge theories possessing singularity-free exact solutions}". The discussion here is closely analogous to the gravitational case 
 \cite{BiswasSiegel,ModestoMoffatNico,BambiMalaModesto2, BambiMalaModesto,calcagnimodesto, koshe1, ModestoGhosts}. 
 In particular the static gauge potential for the exponential form factor $\exp (- \Box/\Lambda^2)$ is for weak fields given approximately by:
 \be
 \Phi_{\rm gauge} ( r ) =  A_0(r) = \con g 
 \frac{{\rm Erf}( \frac{\Lambda r}{2} )}{r} .
 \ee
We used the form factor $\exp (- \Box/\Lambda^2)$ and $D=4$ to end up with a simple analytic solution. However, 
the result is qualitatively the same for the asymptotically polynomial form factor (\ref{VlimitB}), and $\Phi_{\rm gauge} ( r ) = {\rm const}$ for $r=0$.

\subsection{Propagator, unitarity and divergences }
\label{gravitonpropagator}
Splitting the gauge field into a background field (with flat gauge connection) plus a fluctuation, fixing the gauge freedom and computing the quadratic action for the fluctuations, we can invert the kinetic 
operator to get finally the two-point function. This quantity, also known as the propagator in the Fourier space reads, up to gauge dependent components,
\be
&&  
\mathcal{O}^{-1}_{\mu\nu}(k) \!=\!
 \frac{-iV(  k^2/\Lambda^2 )  } {k^2+i\epsilon}  \left( \eta_{\mu\nu} - \frac{k_\mu k_\nu}{k^2} \right) \, ,
 \label{propagator}
\ee
where we used the Feynman prescription (for dealing with poles). The tensorial structure in (\ref{propagator}) is the same of the local Yang-Mills theory, but we see the presence of a new element -- multiplicative form factor $V(z)$. If the function $V^{-1}(z)$ does not have any zeros on the whole complex plane, then the structure of poles in the spectrum is the same as in original two-derivative theory. This can be easily proved in the Coulomb gauge, which is manifestly unitary. Therefore, in the spectrum we have exactly the same modes as in two-derivative theories. In this way we have achieved unitarity, but the dynamics is modified from the simple two-derivative to a super-renormalizable one with higher-derivatives.
Despite that in the UV regime we recover polynomial higher-derivative theory, the analysis of tree-level spectrum still gives us a unitary theory without ghosts, because the renormalizability is due to the behaviour of the theory in the very UV limit, while unitarity is influenced by the behaviour at any energy scale.

In the high energy regime (UV), 
the propagator in momentum space 
schematically scales as 
\be
\mathcal{O}^{-1}(k) \sim k^{-(2 \gamma + D-2) } \,. 
\label{OV} 
\ee
The vertices of the theory can be collected in different sets, that may involve 
or not the entire function $\exp H(z)$. 
However, to find a bound on quantum divergences it is sufficient to concentrate on
the polynomial operators with the high energy leading behaviour in the momenta $k$ \cite{kuzmin, Tombo}.  
These operators scale as the propagator, they cannot have higher power of momentum $k$ in the scaling, in order not to break the renormalizability of the theory. The consideration of them gives the following 
upper bound on the superficial degree of divergence of any graph  \cite{Tombo,modesto,Anselmi0},
\be
\omega(G)\leq DL+(V-I)(2 \gamma +D)-E\,.
\label{omegaG1}
\ee
This bound holds in any spacetime, of even or odd dimensionality. 
In (\ref{omegaG1}) $V$ is the number of vertices, $I$ the number of internal lines, 
$L$ the number of loops, and $E$ is the number of external legs for the graph $G$. 
After plugging the topological relation $I -V= L -1$ in (\ref{omegaG1})  
we get the following simplification:
\be 
&&
\omega(G) \leq  D - 2 \gamma  (L - 1)-E    \,  .
\label{even}
\ee

We comment on the situation in odd dimensions in the next section.
Thus, if in even dimensions $\gamma > (D-E)/2$, in the theory only 1-loop divergences survive.  
Therefore, 
the theory is one-loop super-renormalizable \cite{Krasnikov, Tombo, Efimov, Moffat3,corni1}
and only a finite number of operators of energy dimensions up to $M^D$ 
has to be included in the action to absorb all perturbative divergences. 
In a $D-$dimensional spacetime the renormalizable gauge 
theory includes all the operators up to energy dimension $M^D$, and schematically reads
\be
 \mathcal{L}_D = - \frac{1}{4 g^2}  {\rm tr} \con \left[ {\bf F}^2 + {\bf F}^3 + {\bf F} \,  {\cal D}^2 \, {\bf F} + \dots + {\bf F}^{D/2} \right] . 
\ee

In gauge theory the scaling of vertices originating from kinetic terms of the type 
${\bf F}({\cal D}^2)^{\gamma+ (D-4)/2}{\bf F}$ is lower than the one seen in the inverse propagator 
$k^{2 \gamma +D-2}$. This is because when computing variational derivatives with respect to the dimensionful gauge potentials (to get higher point functions) we decrease the energy dimension of the result. Hence the number of remaining partial derivatives, when we put the variational derivative on the flat connection background, must be necessarily smaller. This means that we have a smaller power of momentum, when the 3-leg (or higher leg) vertex is written in momentum space. We get the maximal scaling for the gluons' 3-vertex, and it is with the exponent $2\gamma+D-3$. In this way we can put an upper bound on the degree of divergence for higher-derivative gauge theories even with a little excess. Again, for higher-derivative gauge theories and $\gamma>(D-E)/2$ we have one-loop super-renormalizability. For the minimal choice $E=2$ (because the tadpole diagram vanishes) we have $\gamma>(D-2)/2$.  

\subsection{Finite gauge theories in odd and even dimensions}
In {\em odd number of dimensions} we can easily show that the theory is finite without need of gauge potential ${\bf V}_{ g}$ because in dimensional regularization scheme (DIMREG) {\em there are no divergences at one-loop
and the theory is automatically finite}. The reason is of dimensional nature. In odd dimension the energy dimension of possible one-loop 
counterterms needed to absorb logarithmic divergences can be only odd. However, 
at one-loop such counterterms cannot be constructed in DIMREG scheme and having at our disposal only Lorentz invariant (and gauge-covariant) building blocks that always have energy dimension two. By elementary building blocks we mean here field strengths or gauge-covariant box operators, or even number of covariant derivatives (even number is necessary here to be able to contract all indices). For details we refer the reader to original papers \cite{modesto}. 

In {\em even dimensions} we for simplicity consider the polynomial $p(z)$ to be a monomial,
$p_\gamma(z)= \omega \, z^{\gamma + \frac{D}{2} - 2}$ ($\omega$ is a positive real parameter). 
In this minimal setup the monomial in UV gives precisely the highest derivative term of the form ${\rm tr}\left(\con{\bf F}({\cal D}^2_{\Lambda})^{\gamma} {\bf F}\right)$ (in $D=4)$. There is only one possible way how to take trace over group indices here, and terms with derivatives can be reduced to those with gauge-covariant boxes only by exploiting Bianchi identities in gauge theory. These latter terms take the explicit form $F_{\mu\nu}^a ({\cal D}^2_{\Lambda})^{\gamma} F^{\mu\nu}_a$. In four dimensions there is an RG running of only one coupling constant. The contribution to the beta function of the YM coupling constant from this quadratic term is actually a dimensionless constant (independent of the frontal coefficient of the highest derivative term), which has been computed in \cite{TesiPiva} using Feynman diagrams. This number can be cancelled by a contribution coming from a quartic (in field strengths) gauge killer of the form 
\be
-\frac{s_g}{4g^2}{\rm tr}\left( {\bf F}^2({\cal D}^2_{\Lambda})^{\gamma-2} {\bf F}^2 \right)
\ee
(here there are several possibilities of taking traces). The contribution to the beta function is linear in the parameter $s_g$ and hence the latter one can be adjusted to make the total beta function vanish.

The action of the finite quantum theory may take the following compact form 
(for the choice $\gamma=3$ the general derivative structure is explicit in $D=4$): 
\be
\label{fingaugeth}
&& \mathcal{L}_{\rm fin,\, gauge} = -\frac{1}{4g^2} {\rm tr}\con\Big[ \underbrace{{\bf F}e^{H({\cal D}_\Lambda^2)}
{\bf F} + s_g {\bf F}^2({\cal D}^2_{\Lambda}) {\bf F}^2}_{\mbox{minimal finite theory}} \nonumber \\
&& + \sum_i \sum_{j>2}^{5} \sum_{k=0}^{{5}-j} c^{(j,k)}_i \left(({\cal D}^2_{\Lambda})^k {\bf F}^j\right)_i\Big]\,,
\ee
where $c^{(j,k)}_i$ are some constant coefficients. The beta function can succesfully killed by the last operator in the first line above. The last terms in the formula \eqref{fingaugeth} have been written in a compact index-less notation and the index $i$ counts all possible contractions of Lorentz and group indices.

\section{The finite theory in $D=4$}
As extensively motivated in the previous section the minimal nonlocal gauge theory 
in $D=4$ candidate to be scale-invariant (finite) at quantum level is: 
\be \!\! 
\mathcal{L}_{\rm fin,\, gauge } = -\frac{\alpha}{4}{\rm tr}\Big[ {\bf F}e^{H({\cal D}_\Lambda^2)}
{\bf F} + s_g {\bf F}^2({\cal D}^2_{\Lambda}) {}^{\gamma-2} {\bf F}^2 \Big]  \, , 
\label{minimalGT}
\ee
where the function $H(z)$ is given in \eqref{VlimitB}.
We here evaluate the contribution to the beta function $\beta_{\alpha}^{(s_g)}$  
from the two following independent  killer operators quartic in the field strength\footnote{It is worth noting that if we choose the gauge group 
$G=SU(N)$ and in the adjoint representation, it holds 
\be
\mathrm{tr}(T^aT^bT^cT^d)=\delta^{ab}\delta^{cd}+\delta^{ad}\delta^{bc} \, , 
\label{Muta} 
\ee
Therefore, the killers we have considered exhaust all the possible operators we can construct, regarding the structure in the internal indices. On top of this we have the freedom of using different contractions of Lorentz indices and covariant derivatives in the expressions for quartic killers. 
Indeed, if we plug the formula above (\ref{Muta}) 
in the following general Lagrangian
\be
{\cal L}_{\rm killer}=-\frac{s_g}{4g^2}\mathrm{tr}\Big[{\bf F}_{\mu\nu} {\bf F}^{\mu\nu} ({\cal D}_{\Lambda}^2)^{\gamma-2} {\bf F}_{\rho \sigma} {\bf F}^{\rho\sigma}\Big] , 
\ee
we get the sum of the two killers (\ref{killer1}) and (\ref{killer2}) 
with the same front coefficient. 
}
\be
&& 1. \,\,\,  -\frac{s_g}{4g^2} F^a_{\mu\nu}F^{\mu\nu}_a \Box_{\Lambda}^{\gamma-2} F^{b}_{\rho\sigma}F^{\rho\sigma}_b
\label{killer1} \,, \\
&& 2. \,\,\, - \frac{s_g}{4g^2} F^{a}_{\mu\nu}F_b^{\mu\nu}({\cal D}^2_{\Lambda})^{\gamma-2}F^{b}_{\rho\sigma}F_a^{\rho\sigma} \, .
\label{killer2}
\ee
All details of the computation are not included in this letter because they are very cumbersome, but the results are:
\be
&& 1. \,\,\,\,  \beta_{\alpha}^{(s_g)}=\frac{s_g}{2\pi^2\omega}, 
\\
&& 2. \,\,\, \, \beta_{\alpha}^{(s_g)}=\frac{s_g}{4\pi^2\omega}(1+N_G),
\ee
where $N_G$ is the number of generators of the Lie group.

These \con results have been checked using two different techniques:  the method of Feynman diagrams and the Barvinsky-Vilkovisky trace technology \con \cite{GBV}. 


The computation has been done  
for the nonlocal theory with general polynomial asymptotic behaviour $p_\gamma(z)$ of degree $\gamma$.
By choosing the monomial $p_\gamma(z)= \omega \, z^\gamma$ 
the prototype kinetic term used to evaluated the beta function reads 
\be
 && \hspace{-1.4cm}\mathcal{L}_{\rm fin,\, \rm kin. gauge} 
=  -\frac{1}{4g^2} F^a_{\mu\nu} \left( 1+  \omega \, ({\cal D}^2_{\Lambda})^{\gamma} \right) 
F_a^{\mu\nu} \, . 
\label{actioncomp}
\ee
As already explained all the other contributions of the form factor fall off exponentially in the UV and
do not contribute to the divergent part of the quantum action. 
To fix our conventions, we can read the beta function from the counterterm operator, namely 
\be
\hspace{0.1cm}
 \mathcal{L}_{\rm ct} := - \frac{\alpha}{4}(Z_\alpha - 1) \, F^{\mu\nu}_a F_{\mu\nu}^a = -  \mathcal{L}_{\rm div} = 
-  \frac{1}{\epsilon} \beta_\alpha \, F_a^{\mu\nu}F^a_{\mu\nu}  . \nonumber 
\ee
\con 
By using the Batalin-Vilkovisky formalism \cite{bata} it is possible to prove that for the theory \eqref{minimalGT} there is no wave-function renormalization for the gauge field 
$A_{\mu}^a$. 
We have only renormalization of the gauge coupling constant. 
The contribution to the beta function $\beta_{\alpha}^{(\gamma)}$ due to the nonlocal kinetic term was obtained in \cite{TesiPiva}, namely 
\be
\beta_{\alpha}^{(\gamma)}= 
 -\frac{(5+3 \gamma +12 \gamma^2)}{192\pi^2}C_2(G) \, , \quad    \gamma \geq 2 \, , 
\label{betas}
\ee 
where $C_2(G)$ is the quadratic Casimir of the gauge group $G$.  
  By imposing the following condition for scale invariance 
\be
\beta^{(\gamma)}_{\alpha}+\beta_{\alpha}^{(s_g)}=0 ,
\ee
we can find the special value of the coefficient $s_g^*$ that kills the beta function. 
Using for example the first killer (\ref{killer1}) we get
\be
s_g^*=-2\pi^2  \omega \beta_{\alpha}^{(\gamma)} ,
\label{adjusting}
\ee
and the Lagrangian for a finite nonlocal gauge theory in four dimensions can be explicitely written
\be
&& \mathcal{L}_{\rm fin,\, gauge } = -\frac{\alpha}{4} \Big[ F_{\mu\nu}^ae^{H({\cal D}_\Lambda^2)}
F^{\mu\nu}_a 
\label{finita} \\
&& +  \omega \frac{(5+3 \gamma+12 \gamma^2)}{96}C_2(G) F_{\mu\nu}^a F^{\mu\nu}_a({\cal D}_{\Lambda}^2)^{\gamma-2} F^b_{\rho \sigma} F_b^{\rho\sigma}  \Big] \nonumber 
\ee
where we assumed $\gamma \geq 2$.

It is possible to kill the beta function also in nonlocal theories, where we have Abelian symmetry groups. For concreteness we can study the one-loop beta function of QED $\beta_e=  e^3/12\pi^2$
 for electric charge $e$. In terms of the inverse coupling $\alpha$ this function is expressed as 
 $\beta_\alpha= - 1/6\pi^2$, 
 which is a constant and gives logarithmic scaling with the energy for the coupling constant $\alpha$. Since pure two-derivative QED is a free theory, then the running comes entirely from quantum effects of charged matter. Here we assume one species of charged fermions coupled minimally to photon field. If we extend 
 QED to the nonlocal version \eqref{gauge} with killer operator (\ref{killer1}) and we replace 
\be
s_g^*=-2\pi^2 \omega \beta_{\alpha}^{(\gamma)} = \frac{ \omega }{3}
\label{adjustingQED}
\ee
 in \eqref{minimalGT}, then the theory is completely finite regardless of the parameter $\gamma$. It is important to notice that even in the Abelian case the killer operator has crucial impact on the beta function because it contains photon self-interactions. In this way we solve the problem of Landau pole for the running of the electric charge in the UV regime of QED. The same can be repeated for any gauge theory coupled to matter, provided that in the matter sector we do not have self-interactions and the coupling to gauge fields is minimal \cite{TesiPiva}.

We want to comment on what we can achieve if we stick to one-loop super-renormalizable gauge theories without attempts to make them finite. 
The final result \eqref{betas} \con highlights a universal Landau pole issue in the UV regime for the running coupling constant $g(\mu)$ (where $\mu$ is the renormalization scale).
This is true for any value of the integer $\gamma \geq2$, when \con we do not introduce any potential 
${\bf V}_g$ with killer operators. The sign of the beta function is negative because the discriminant 
$\Delta<0$ of the quadratic polynomial in $\gamma$ in \eqref{betas}. 
 For the particular choice (\ref{adjusting}) the theory  \eqref{minimalGT} is one-loop finite, but if the front coefficient $s_g$ has a bigger value than in (\ref{adjusting}) then we enter the regime in which  the UV asymptotic freedom is achieved. 
We here summarize the three possible scenarios for the value of the $s_g$
\be
s_g \left\{
\begin{array}{lr}
  < \omega\frac{(5+3 \gamma+12 \gamma^2)}{96}C_2(G) \, ,   \quad \mbox{Landau pole} , \\ 
  \\
 =   \omega\frac{(5+3 \gamma+12 \gamma^2)}{96}C_2(G)  \equiv s_g^*  \, , \quad \mbox{finiteness}   , \\
 \\
 > \omega \frac{(5+3 \gamma +12 \gamma^2)}{96}C_2(G) \, , \quad   \mbox{asymptotic freedom} .
\end{array}
\nonumber 
\right.
\ee 
However, 
in weakly nonlocal higher-derivative theories we must read out the poles  
from the quantum effective action and not only from the beta functions of the couplings in the theory. 
In particular, in the case of the theory \eqref{gauge} the one-loop dressed propagator is devoid of any pole because its UV asymptotic behaviour 
is entirely due to the form factor $\exp H(z)$ \cite{Tombo}, namely, up to the tensorial structure, 
\be
- i \frac{e^{- H(k^2) }}{k^2 \left( 1 + \beta_{\alpha} \, e^{- H( k^2)  } \log ( k^2/\mu_0^2)  \right)  }.
\ee
Moreover, as a particular feature 
of the super-renormalizable theory, when \con $s_g=0$ or $s_g < s_g^*$, 
$\beta_{\alpha}$
 is negative, signifying that at low energy the theory is weakly coupled. In consequence \con
we do not have any pole in the dressed propagator in the UV nor do we have any problem in the IR 
as opposite to the local theory. 

In local two-derivative theories we usually have a UV Landau pole or an IR singularity of RG flow, so 
(as for example in QED) the theory is weakly coupled in the IR (without confinement), but it becomes 
non-perturbative in the UV.
In QCD we have the reverse, the theory is asymptotically free in the UV where it is perturbative, but a 
singularity of the RG flow  
manifests itself  in the IR indicating confinement. In the case of two-derivative local theories the singularities of the flow have direct realization as the poles in the effective propagator read from the quantum action. This is not true anymore when higher-derivatives are included. 
In the theory (\ref{minimalGT}) for $s_g < s_g^*$, the minus sign of the beta function, which usually gives rise to a UV Landau pole, is innocuous because the form factor washes away the $\log (k^2)$ contributions to the dressed propagator in the UV and there is no possibility for appearance of a new real pole in it.  
On the other hand, in the IR the analytic form factor does not play any role and  
there is no pole because the beta function is negative. The outcome is a theory perturbative in both the UV and in the IR regime. 
Therefore we are left with two possible options.
We can choose completely UV finite (no divergences) nonlocal theories or super-renormalizable nonlocal theories with negative beta functions ($\beta_\alpha$) and hence without any singularities in asymptotic behaviours of the couplings. The second option seems to be very 
appealing in models that attempt to realize a unification of all coupling constants.

\section{Conclusions}
We have explicitly evaluated the one-loop exact beta function for the weakly nonlocal gauge theory
recently proposed in \cite{modestoLeslaw}. The higher-derivative structure or 
quasi-polynomiality of the action implies that the theory is
super-renormalizable, and in particular only one-loop divergences survive in any dimension. 
Once a potential, at least cubic in the field strengths, is switched on, it is always possible to make the theory finite.
We evaluated the beta function for the special case of $D=4$, but the result can be generalized 
to any dimension where a careful selection of the killer operators should be done. 

In short, in this paper 
{\em we have explicitly shown how to construct a finite theory for 
 gauge bosons in $D=4$} (\ref{finita}). We have considered both cases of Abelian and non-Abelian gauge symmetry groups. 
The super-renormalizable structure does not change if we add a general extra matter sector 
that does not exhibit self-interactions.

The minimal nonlocal theory without any killer operator shows a Landau pole 
for the running coupling constant, regardless of the special asymptotic polynomial structure. This is a universal property shared at least by all the unitary and weakly nonlocal gauge theories with asymptotic polynomial behaviour in the UV regime. However, the one-loop dressed propagator does not show any Landau pole
in the UV regime because the propagator is dominated by the nonlocal form factor and it is the nonlocal  operator to control the high energy physics. Moreover, 
we do not have any pole even in the IR, as opposite to the local theory, exactly because the universal 
negative sign of the beta function. 
The outcome is a theory well defined at perturbative level in both the IR and the UV regime.
The same result is achieved in the presence of sufficiently weakly coupled killer operators.


In this paper we mostly considered pure gauge theories, but here we can achieve asymptotic freedom 
regardless of the number of fermionic fields, because it is the interaction between gauge bosons, due to the killer operators, that makes the theory asymptotically free.

The generalization to extra dimensions is straightforward.
In particular, the theory is finite in odd dimension without the need to introduce any killer operator, as a mere consequence of dimensional regularization. 
The results can also be reproduced in cut-off regularization making use of Pauli-Villars operators \cite{Anselmi2}.

{\em Acknowledgements ---}
We are grateful to D. Anselmi for very useful discussions in quantum field theory.

\end{document}